# Dual-Polarization Second Harmonic Generation Interferometry for Imaging Antiparallel Domains and Stacking Angles of 2D Heterocrystals

Juseung Oh[1*], Wontaek Kim[1], Gyouil Jeong[1], Yeri Lee[1], Jihun Kim[1], Hyeongjoon Kim[2,3], Hyeon Suk Shin[2,4,5] and Sunmin Ryu[1,6*]

[1]Department of Chemistry, Pohang University of Science and Technology (POSTECH), Pohang, Gyeongbuk 37673, Korea

[2]Center for 2D Quantum Heterostructures, Institute of Basic Science (IBS), Suwon 16419, Korea

[3]Department of Chemistry, Ulsan National Institute of Science and Technology (UNIST), Ulsan 44919, Korea

[4]Department of Chemistry, Sungkyunkwan University (SKKU), Suwon 16419, Republic of Korea

[5]Department of Energy, Sungkyunkwan University (SKKU), Suwon 16419, Republic of Korea

[6]Institute for Convergence Research and Education in Advanced Technology (I-CREATE), Yonsei University, Seoul 03722, Korea

*E-mail: juseungoh@postech.ac.kr, sunryu@postech.ac.kr

## ABSTRACT

Optical second-harmonic generation (SHG) enables orientational polarimetry for crystallographic analysis and domain imaging of various materials. However, conventional SHG intensity polarimetry, which neglects phase information, fails to resolve antiparallel domains and to describe two-dimensional heterostructures, which represent a new class of van der Waals-bound composite crystals. In this work, we report dual-polarization spectral phase interferometry (DP-SPI) and establish a generalized SHG superposition model that incorporates the observables of DP-SPI. Antiparallel domains of monolayer transition metal dichalcogenides (TMDs) were successfully imaged with distinction, validating the interferometric polarimetry. From DP interferograms of TMD heterobilayers, the orientation of each layer could be determined, enabling layer-resolved probing. By employing the superposition model, we also demonstrate the photonic design and fabrication of ternary TMD heterostructures for circularly polarized SHG. These methods, providing comprehensive SHG measurements and theoretical descriptions, can be extended to heterostructures consisting of more than two constituent layers and are not limited to TMDs or 2D materials.

KEYWORDS: second-harmonic generation (SHG), polarimetry, spectral phase interferometry, two-dimensional (2D) heterostructures, transition metal dichalcogenides (TMDs)

## Introduction

Reduced dimension and the consequent modulation of dielectric screening[1] lead to strong and rich light-matter interactions in two-dimensional (2D) crystals ranging from tightly bound excitons[2-4] and their polaritons[5] to valley degree of freedom[6, 7] and photonic topological states[8]. Crystalline 2D heterostructures formed via van der Waals (vdW) interactions with controlled stacking angle also exhibit novel material properties: Superconductivity[9] and correlated insulator[10] emerge in bilayer graphene stacked at a specific angle. Interlayer excitons[11, 12] are efficiently formed owing to ultrafast[13] charge separation in Type-II heterostructures with their lifetime[14] and transport[15] governed by the stacking angle. Moiré superlattices dictated by the stacking angle endow another dimension to control the electronic structures[16] and associated photophysical phenomena[17]. These emergent properties and their dependence on stacking angle indicate that efficient and reliable crystallographic characterization of 2D heterostructures is crucial for additional discovery of novel principles and applications.

Optical second-harmonic generation (SHG) has provided a polarimetric method to characterize the crystalline structures and crystallographic orientation of various low-dimensional systems ranging from interfacial monolayers[18, 19] and muscle proteins[20] to molecular nanocrystals[21], quantum dots[22] and nanowires[23, 24]. More recently, it has been widely applied to 2D materials including hexagonal boron nitride (hBN)[25-28] and transition metal dichalcogenides (TMDs)[25, 29-32]. The strengths of SHG polarimetry are manifold in general and more notable for 2D systems: First of all, SHG generally provides anisotropic responses even when absorption, emission and Raman scattering are isotropic and unusable for orientational polarimetry. This arises because the nonlinear susceptibility responsible for SHG is a third-rank tensor and more complex than the linear term[33]. For atom-thin 2D materials, SHG is not restricted by the phase-matching requirement[31], is sufficiently strong in even single-layer (1L) hBN[25] and is orders of magnitude enhanced by excitonic resonances in 2D semiconductors including TMDs[29, 34-36]. SHG spectroscopy also has high spatial resolution limited by optical diffraction and can reach a super-resolution limit in a tip-enhanced scheme[37-39]. Lastly, it is a fast and non-destructive method not requiring additional treatment of samples unlike electron diffraction or atomic-resolution probes using electron tunneling or transmission.

Despite its successful application to various 2D materials including continuously grown 1L $MoS_2$[30] and hBN[26, 28], conventional intensity-based polarimetry has revealed critical limitations by neglecting the phase information of coherent SHG signals. Most notably, the method fails to differentiate two antiparallel domains, which, for example, can be formed by 60° rotation for 1L TMDs and hBN. This capability is essential for characterizing large-area 2D materials typically grown on centrosymmetric catalytic surfaces, which do not preferentially favor either domain. Second, the intensity polarimetry is incapable of characterizing artificially stacked heterobilayers, which require the amplitude and phase information on the two SH fields that contribute to the vectorial superposition[32]. In particular, phase delays in SHG process[31] can be significant near electronic resonances and are therefore crucial for interpreting signals detected in the far field. Notably, it is still possible to perform limited orientational polarimetry using conventional schemes when inter-material SHG phase difference is negligible[40] or when each layer can be exclusively probed via excitonic resonances[41]. Also note that stacking angle of 30°, generating orthogonally polarized SH fields, allows independent probing of each layer in TMD heterobilayers.[42, 43] Nevertheless, a generalized superposition model and phase-sensitive polarimetric method are required to characterize the SH fields and crystallographic orientations of heterostructures.

In this work, we report an interferometric SHG polarimetry technique and a generalized superposition model that overcome the aforementioned limitations. Using newly developed dual-polarization spectral phase interferometry (DP-SPI), we demonstrate unprecedented orientational imaging that reveals antiparallel domains of 1L $WSe_2$. By devising a matrix representation of the superposition model, we show that two orthogonal interferograms from DP-SPI measurements enable layer-resolved SHG analysis of heterobilayers and determine the orientation of each constituent layer in $MoS_2/WS_2$ heterobilayer and thus its stacking angle. As a proof-of-concept demonstration of nonlinear photonic design enabled by the generalized superposition model, we fabricated ternary heterostructures and verified the generation of circularly polarized SHG signals.

## Results and Discussion

***Dual-polarization spectral phase interferometry.*** The SHG signals from TMD bilayers can be described as the superposition of those from individual layers[31]. As shown below, the SHG responses of individual layers can then be extracted from the overall SHG intensity and phase values when these are decomposed into two orthogonal polarizations. For this purpose, we devised the DP-SPI technique that measures the SHG response of the whole system in the two orthogonal polarizations. The DP interferograms were then analyzed with the newly developed generalized superposition model, which relates them to the SHG responses of individual layers. Below, we present the machinery of DP-SPI and the superposition model along with their representative applications.

In the DP-SPI setup shown in Fig. 1a, the fundamental pulse (denoted as 1ω) polarized along the y axis (Fig. 1b, top) passed through a z-cut α-quartz crystal, and a small fraction of the pulse was converted to an SHG pulse ($2\omega_{\mathrm{LO}}$) that served as a local oscillator (LO)[31]. The quartz crystal was azimuthally oriented to make the x (⊥) and y (∥) components of $2\omega_{\mathrm{LO}}$ equal in intensity. The remaining intensity of the fundamental pulse interacted with the sample to generate another SHG pulse ($2\omega_{\mathrm{sam}}$). Each of the two SHG signals was split by a Wollaston prism into two orthogonal polarizations (∥ and ⊥). Then, the two pulses ($2\omega_{\mathrm{sam}}$ and $2\omega_{\mathrm{LO}}$), temporally broadened in the spectrometer, generated two interferograms (Fig. 1c) of mutually orthogonal polarizations at the CCD. For non-interferometric polarization-resolved SHG measurements, a long-pass filter was inserted right before the objective lens to block $2\omega_{\mathrm{LO}}$. Note that the interference scheme for homobilayers by S. Psilodimitrakopoulos et al.[44] is conceptually similar in that half of the bilayer generates LO.

In Fig. 1c, we present the DP interferograms $\tilde{I}^{\parallel(\perp)}(\omega)$ of 1L $MoS_2$ obtained for three different orientational angles ($\theta^o$): $\theta^o$ is defined as the azimuthal angle of the armchair direction ($\overrightarrow{AC}$) measured counter-clockwise from the +y axis. To be consistent with the three-fold rotational symmetry of 1L TMD, $\overrightarrow{AC}$ was defined as a vector pointing from one type of atom to its diagonal atom in a hexagonal unit cell (Fig. 1b) with $\theta^o$ varying from 0 to 120°. Then, $\tilde{I}^{\parallel(\perp)}(\omega)$ can be described as follows[31]:

$$\tilde{I}^{\parallel(\perp)}(\omega) \propto \left\{ {E_{sam}^{\parallel(\perp)}}^2 + {E_{LO}^{\parallel(\perp)}}^2 + 2E_{sam}^{\parallel(\perp)}E_{LO}^{\parallel(\perp)}\cos(\tau\omega - 2\omega_0\tau + \varphi) \right\} G(\omega - 2\omega_0) \qquad (1)$$

, where $E_{sam}^{\parallel(\perp)}$ and $E_{LO}^{\parallel(\perp)}$ are the polarized SH field amplitude of $2\omega_{sam}$ and $2\omega_{LO}$, respectively. $G$, $\omega_0$, $\tau$ and $\varphi$ are the Gaussian envelope function, the central frequency of the fundamental pulse, the time delay and the phase difference between the two SHG pulses, respectively. Note that $G$ corresponds to the line shape obtained in a conventional SHG spectroscopy. Equation 1 reveals that a spectral oscillation with a period of $2\pi/\tau$ sits on top of two constant SH contributions from the sample and the local oscillator. We note a complementary intensity variation in $\tilde{I}^{\parallel}$ and $\tilde{I}^{\perp}$ with a period of $60°$ (Fig. 1c), which is governed by the $D_{3h}^1$ symmetry of 1L TMD as follows[25]: $I^{\parallel} \equiv {E_{sam}^{\parallel}}^2 \propto \cos^2 3\theta^o$, $I^{\perp} \equiv {E_{sam}^{\perp}}^2 \propto \sin^2 3\theta^o$.

***Interferometric orientational polarimetry.*** Before discussing the interference of two SH fields generated in heterobilayers, we will show what information interferometric SHG spectroscopy can bring regarding the orientational polarimetry of 1L TMDs. The so-called orientational polarimetry by SHG intensity has been widely used to determine the crystallographic orientation of 2D TMDs[25, 30, 31, 45, 46], for which linear spectroscopies are of no use because of their in-plane isotropic responses. However, the method has a few limitations. First, as $\theta^o = \pm\frac{1}{3}\tan^{-1}\left\{\left(\frac{I^{\perp}}{I^{\parallel}}\right)^{1/2}\right\}$, two orientations of the same angular deviation are generated as a possible solution for a given measurement except when $\theta^o =$ zero. The correct one can be found by making another measurement after rotating the samples slightly because it is a matter of rotational direction, clockwise vs. counterclockwise. Second, the above trigonometric relations further reveal that a sample rotation by $60°$ maintains the SHG intensity despite the inequivalence of the two orientations. This indeterminacy means that the method cannot differentiate two antiparallel domains of 1L TMDs[30] for example. Noting that such rotation leads to the reversal of SH fields or 180° change in φ[31], one can devise an interferometric polarimetry approach that can remove the limitation inherent in the intensity polarimetry and directly distinguish between two antiparallel domains. By providing the SHG phase as well as polarization-resolved SHG intensity, DP-SPI satisfies the requirements for the interferometric polarimetry.

As a proof-of-principle demonstration, we performed intensity and interferometric polarimetry measurements for two 1L $WSe_2$ crystals grown in an apparently antiparallel manner (Fig. 1d). The two crystals were almost indistinguishable in the conventional SHG intensity image (Fig. 1e) and spectra obtained for the two polarizations (Fig. 1f). The crystallographic orientation (θ°) for each pixel could be determined from polarized intensity images ($I^{\parallel}$ and $I^{\perp}$, Fig. S1) and the above trigonometric relation: As shown in their θ° image (Fig. 1g), the two crystals could hardly be distinguished. The θ° histogram in Fig. S2a showed that orientational mismatch was only 2.5 ± 0.6°. However, these data do not tell whether they are aligned parallel or antiparallel, as explained above. In Fig. 1h, we show $\tilde{I}^{\parallel}$ obtained from the two crystals: The spectral oscillations are staggered with a half-cycle phase difference (180°). Note that slowly varying components in Equation 1 were removed by Fourier filtering as shown in Fig. S3[31]. The phase (φ) image (Fig. 1i) revealed that they are indeed antiparallel to each other and also confirmed that they are single crystals without any orientational sub-domains. In the phase histogram (Fig. 1j), the two crystals appear as two phase groups with narrow distributions. The resolution of the phase measurements was less than 5°, which was determined using mechanically exfoliated 1L $WS_2$ as a reference. When the phase information was incorporated, the θ° image in Fig. 1k revealed the true orientations of the two crystals. The phase-resolved θ° histogram in Fig. S2b showed that they differ by 54.7°, not 2.5° in their $\overrightarrow{AC}$ orientation. We also note that the interferometric polarimetry can be applied to other material systems including hBN and those with lower symmetries.

***Generalized SHG superposition model of heterobilayers.*** A generalized description of SHG by heterobilayers (Fig. 2a) can be formulated based on the vector superposition model by W. Hsu et al.[32]. For a homobilayer specified with a stacking angle $\theta_S$, they showed that its SHG signals have a maximum (4 times of monolayer) for $\theta_S = 0°$, zero for $\theta_S = 60°$, and intermediate values for other $\theta_S$. One key ingredient missing in the previous scheme is the material-dependent phase delay in SHG[31]. In heterobilayers, constructive or destructive interference is governed by not only $\theta_S$ between the two layers but also the difference in their phase. Because of their phase difference, the SHG signals of heterobilayers are elliptically polarized unless the SHG fields are aligned, either parallel or antiparallel[31].

To validate and characterize the generalized SH superposition model for heterobilayers, we investigated how their SHG intensity and phase behave as a function of their orientation. In this study, we fabricated TMD heterobilayers consisting of 1L $WS_2$ and $MoS_2$ on amorphous quartz substrates using the dry transfer method.[47] The morphology and structural quality of the samples were characterized with AFM imaging (Fig. 2b), Raman (Fig. S4a) and photoluminescence spectra (Fig. S4b). The height image in Fig. 2b shows a heterostructure (HS) area of ~30 $\mu m^2$, where $MoS_2$ is stacked on $WS_2$ (denoted as $MoS_2/WS_2$). Their stacking angle, judged from their polar graphs for $I^{\parallel}$ in Fig. S5a, was 32.9 ± 0.2°. Despite its six-fold symmetry, however, the response from the HS area shown in Fig. 2c (blue circles) is unusual in that its minima are far from zero. Given the presence of a polarizer in front of the detector, these nonzero minima indicate that the SHG signals from HS is not plane-polarized but elliptically polarized[31].

The SHG signals from the HS area (Fig. 1b, bottom) can be described using a dual-polarization superposition model depicted in Fig. 2a, where two SH fields of orthogonal polarizations ($E^{\parallel}$ and $E^{\perp}$) are given as follows:

$$E^{\parallel} = \sum_{j=1}^{2} \alpha_j \cdot \cos 3\left(\theta_j^{o} - \theta\right) \cdot e^{i(2\omega t - \varphi_j)} \qquad (2)$$

$$E^{\perp} = \sum_{j=1}^{2} \alpha_j \cdot \sin 3\left(\theta_j^{o} - \theta\right) \cdot e^{i(2\omega t - \varphi_j)} \qquad (3)$$

, where $\alpha_j$ and $\varphi_j$ are the SH amplitude and phase of $j^{th}$ layer, respectively. The difference between $\theta_j^{o}$ and $\theta$ specifies the orientational angle of $j^{th}$ layer when HS is rotated clockwise by $\theta$ from its original orientation (Fig. 1b, bottom). Using Equation 2, $I^{\parallel}$ was calculated and shown in the orange dashed line in Fig. 2c. Note that all the quantities in Equation 2 were obtained from the two monolayer areas. The good match between the simulation and experimental data validates the generalized superposition model.

Equations 2 and 3 also enable intuitive understanding of the nature of the SHG signals from the HS area. The SH field in each polarization consists of two sinusoidal components with distinct amplitudes and phases. Their superposition is mathematically equivalent to a single sinusoidal wave characterized by a specific amplitude and phase. To analyze the relation of these waves, we present the polarization-resolved interferograms of 1L $MoS_2$ ($\tilde{I}^{\parallel}_{Mo}$), 1L $WS_2$ ($\tilde{I}^{\parallel}_{W}$) and HS ($\tilde{I}^{\parallel}_{HS}$) areas in Fig. 2d (see Fig. S6 for cross-configuration data). The 1L area exhibited a constant phase value for a $\theta$ interval of 60° and a flipped phase for the next 60° interval. In contrast, HS showed gradual phase changes with the same periodicity. The phase values extracted from the interferograms (Fig. 2e for $\varphi^{\parallel}_{Mo}$ and $\varphi^{\perp}_{Mo}$; Fig. 2f for $\varphi^{\parallel}_{W}$ and $\varphi^{\perp}_{W}$) revealed intriguing features: i) phase flips in the two polarizations with an offset of 30° in $\theta$ and ii) inter-material phase difference ($\varphi^{\parallel}_{Mo}$ vs. $\varphi^{\parallel}_{W}$). Furthermore, the wavy change of $\varphi^{\parallel}_{HS}$ in Fig. 2g could be well predicted (orange dashed line) using the superposition model, as will be explained below.

The $\theta$-dependent phase behaviors of 1L and HS areas can be explained graphically using the DP superposition model. Because the SH field of 1L $MoS_2$, $\boldsymbol{E}_{\mathbf{Mo}}$ in Fig. 2h, rotates counterclockwise by $3\theta$ with samples being rotated clockwise by $\theta$, the x ($\perp$) or y ($\parallel$) component of $\boldsymbol{E}_{\mathbf{Mo}}$ changes its sign every 60°. Then, the phase flip in Fig. 2e occurs for the parallel (cross) polarization configuration when the zigzag (armchair) directions of 1L $MoS_2$ are aligned along the y axis. 1L $WS_2$ also showed an identical alternation but with significant offsets along both axes as shown in Fig. 2f. The horizontal offset is due to the stacking angle ($\theta_S$), which places $\boldsymbol{E}_{\mathbf{W}}$ after $\boldsymbol{E}_{\mathbf{Mo}}$ by $3\theta_S$ (Fig. 2h). The vertical offset corresponds to the difference in the phase delays between the two materials ($\Delta\varphi = 61^{\circ}$ at 400 nm)[31]. The change in $\varphi^{\parallel}_{HS}$ can be described on the complex plane, where $E^{\parallel}_{HS}$ spanned by $E^{\parallel}_{Mo}$ and $E^{\parallel}_{W}$ rotates counterclockwise at an angular frequency of $2\omega$ with a phase ($\varphi^{\parallel}_{HS}$) as depicted in Fig. 2i. Notably, $\varphi^{\parallel}_{HS}$ varies when the sample is rotated because $E^{\parallel}_{Mo}$ and $E^{\parallel}_{W}$ depend on the orientational angle (Fig. 2h). A similar vector construction for the other polarization ($E^{\perp}_{HS}$) can be made to give $\varphi^{\perp}_{HS}$. The $\theta$-dependence of heterobilayer's phase could be successfully predicted (orange dashed line in Fig. 2g) based on the SHG information of constituent 1Ls using a matrix approach presented below. This enabled the design of vdW-stacked heterocrystals for NLO photonic purposes as will be demonstrated later.

***C-matrix representation of SHG interference.*** Although Equations 2 and 3 are straightforward to understand and provide the above geometric interpretation of SH interference, they are not well suited for factoring out the SH amplitude, phase or orientational information. To address this need and boost the analytical power of the superposition model, we devised its matrix representation, which can be extended to N-layer systems (Supplementary Note A). The two equations can be combined into the following:

$$(\mathrm{E}^{\parallel} \quad \mathrm{E}^{\perp}) = [1 \quad 1] \cdot \boldsymbol{\Phi} \cdot \mathbf{A} \cdot \boldsymbol{\Theta} \cdot \begin{bmatrix} \cos 3\theta & -\sin 3\theta \\ \sin 3\theta & \cos 3\theta \end{bmatrix} \cdot \mathrm{e}^{2i\omega t} \qquad (4)$$

, where $\boldsymbol{\Phi}$, $\mathbf{A}$ and $\boldsymbol{\Theta}$ are the matrices consisting of $\varphi_j$, $\alpha_j$ and $\theta_j$, and respectively defined as follows:

$$\boldsymbol{\Phi} \equiv \begin{bmatrix} \cos\varphi_1 & \cos\varphi_2 \\ -i\sin\varphi_1 & -i\sin\varphi_2 \end{bmatrix} \qquad \mathbf{A} \equiv \begin{bmatrix} \alpha_1 & 0 \\ 0 & \alpha_2 \end{bmatrix} \qquad \boldsymbol{\Theta} \equiv \begin{bmatrix} \cos 3\theta_1^{o} & \sin 3\theta_1^{o} \\ \cos 3\theta_2^{o} & \sin 3\theta_2^{o} \end{bmatrix} \qquad (5)$$

We further define $\boldsymbol{\Phi} \cdot \mathbf{A} \cdot \boldsymbol{\Theta}$ as a new matrix $\mathbf{C}$, which carries all SHG characteristics of the two monolayers as follows:

$$\mathbf{C} = \boldsymbol{\Phi} \cdot \mathbf{A} \cdot \boldsymbol{\Theta} = \begin{bmatrix} \mathrm{C}_{11} & \mathrm{C}_{12} \\ i\mathrm{C}_{21} & i\mathrm{C}_{22} \end{bmatrix} \qquad (6)$$

Because the $\mathbf{C}$-matrix elements can also be evaluated from DP-SPI measurements of the HS areas (Supplementary Note B), Equation 6 connects the experimental observables of heterobilayers with the SHG properties of individual monolayers. With matrix inversion, Equation 6 enables the description of $\boldsymbol{\Phi}$, $\mathbf{A}$ or $\boldsymbol{\Theta}$ in terms of $\mathbf{C}$ and the other matrixes: $\boldsymbol{\Theta} = \mathbf{A}^{-1} \cdot \boldsymbol{\Phi}^{-1} \cdot \mathbf{C}$, for example. This fact is important because it allows one to obtain constituent monolayer's information by probing HS areas. This, so-called layer-resolved probing, will be demonstrated below for the crystallographic orientation ($\boldsymbol{\Theta}$) of $MoS_2/WS_2$.

Using the $\mathbf{C}$ matrix, one can predict the polarized SHG intensities ($\mathrm{I}^{\parallel}$ and $\mathrm{I}^{\perp}$) and phases ($\varphi^{\parallel}$ and $\varphi^{\perp}$) of heterobilayers (Supplementary Note B):

$$\mathrm{I}^{\parallel} \propto \left(\mathrm{C}_{11}{}^2 + \mathrm{C}_{21}{}^2\right)\cos^2 3\theta + \left(\mathrm{C}_{12}{}^2 + \mathrm{C}_{22}{}^2\right)\sin^2 3\theta + (\mathrm{C}_{11}\mathrm{C}_{12} + \mathrm{C}_{21}\mathrm{C}_{22})\sin 6\theta \qquad (7)$$

$$\mathrm{I}^{\perp} \propto \left(\mathrm{C}_{12}{}^2 + \mathrm{C}_{22}{}^2\right)\cos^2 3\theta + \left(\mathrm{C}_{11}{}^2 + \mathrm{C}_{21}{}^2\right)\sin^2 3\theta - (\mathrm{C}_{11}\mathrm{C}_{12} + \mathrm{C}_{21}\mathrm{C}_{22})\sin 6\theta \qquad (8)$$

$$\varphi^{\parallel} = -\tan^{-1}\left(\frac{C_{21} + C_{22}\tan 3\theta}{C_{11} + C_{12}\tan 3\theta}\right) \qquad (9)$$

$$\varphi^{\perp} = -\tan^{-1}\left(\frac{C_{22} - C_{21}\tan 3\theta}{C_{12} - C_{11}\tan 3\theta}\right) \qquad (10)$$

As shown for $I^{\parallel}$ (Fig. 2c) and $\varphi^{\parallel}$ (Fig. 2g), the prediction agreed well with the experimental data, which was also confirmed for the cross configuration (Fig. S7). We also validated the superposition model for other systems: i) $WS_2/MoS_2$, where stacking order was reversed (Fig. S8a), ii) $WS_2/MoS_2$ with $\theta_S = 15.0°$ (Fig. S8d), iii) $WS_2/MoSe_2$ (Fig. S8g) and iv) $WS_2/WSe_2$ (Fig. S8j). Besides validating the model, the agreement indicates that vdW stacking did not affect SHG response of each layer significantly. Note that such non-interacting picture is not valid near excitonic resonances or in high-fluence limit[43].

**Layer-resolved orientational imaging of heterobilayers.** As a proof-of-principle demonstration of layer-resolved probing, we present orientational imaging of each layer in heterobilayers. The intensity polarimetry that is useful for monolayer TMDs fails for heterobilayer systems because of the aforementioned SH interference except for a few limited cases[40, 41]. As shown earlier, the orientational angles of the constituent monolayers ($\mathbf{\Theta}$, or $\theta_1^o$ and $\theta_2^o$) can be obtained from the definition of **C**: $\mathbf{\Theta} = \mathbf{A}^{-1} \cdot \mathbf{\Phi}^{-1} \cdot \mathbf{C}$, which leads to $\theta_1^o$ and $\theta_2^o$ as follows (Supplementary Note C):

$$\theta_1^o = f_{2L}(\varphi_2) \text{ or } 60° - f_{2L}(\varphi_2) \qquad (11)$$

$$\theta_2^o = f_{2L}(\varphi_1) \text{ or } 60° - f_{2L}(\varphi_1) \qquad (12)$$

$$\text{, where } f_{2L}(\varphi) \equiv \left|\frac{1}{3}\tan^{-1}\left(\frac{C_{22}+C_{12}\tan\varphi}{C_{21}+C_{11}\tan\varphi}\right)\right|$$

Equations 11 and 12 indicate that $\theta_1^o$ and $\theta_2^o$ can be found by determining the matrix elements of **C**. As we showed in Supplementary Note B, **C** can be evaluated with DP-SPI, which provides $I^{\parallel}$, $I^{\perp}$, $\varphi^{\parallel}$ and $\varphi^{\perp}$. Notably, it does not require prior information on individual 1Ls or probing 1L areas except for $\varphi_1$ and $\varphi_2$.

For experimental verification, we performed dual-polarization intensity and phase imaging on $MoS_2/WS_2$ shown in Fig. 3a. As shown in Fig. 3b, the image of total intensity ($I^{\parallel} + I^{\perp}$) agrees well with the optical micrograph of the heterobilayer sample (Fig. 3a). Because the SHG signals of $WS_2$ were much larger than those of $MoS_2$, the HS area was largely dominated by the signals of 1L $WS_2$. For each of the image pixels, the **C** and subsequently $\boldsymbol{\Theta}$ matrixes were calculated using $I^{\parallel}$, $I^{\perp}$, $\varphi^{\parallel}$ and $\varphi^{\perp}$ values for the corresponding pixel (see Fig. S9 for their images). In Figs. 3c and 3d, we show the orientational images of $\theta_{W}^{o}$ and $\theta_{Mo}^{o}$, respectively. Between the two solutions of Equation 11 (and 12), the one that satisfies $0 \leq \theta_{j}^{o} < 30°$ was displayed. The $\theta_{W}^{o}$ image in Fig. 3c reveals that the orientational angle of $WS_2$ is essentially equivalent between the 1L and HS areas. Note that the data were corrected for the phase shifts induced by multiple reflections and absorptions as explained later. The $\theta_{W}^{o}$ histograms in Fig. 3e shows that the distribution of the HS area is ~5 times wider than that for 1L, the FWHM of which turned out to be 2.3°. The broader width of the HS area can be attributed to structural irregularities including bubble-like features at the interface (see Fig. 2b for morphology). A similar agreement could be found for $MoS_2$ ($\theta_{Mo}^{o}$) as shown in Figs. 3d and 3f. The average stacking angle ($\theta_S = |\theta_{W}^{o} - \theta_{Mo}^{o}|$ or $|60° - \theta_{W}^{o} - \theta_{Mo}^{o}|$) obtained from the histograms in Figs. 3e and 3f was (9.8 or 34.3) ± 5.5°, the latter of which was consistent with $\theta_S$ determined by the intensity polarimetry of 1L areas (Fig. S5a). This example demonstrates the layer-resolved probing capability of DP-SPI when combined with the **C**-matrix analysis.

***Non-interferometric determination of SHG phase.*** A proper description of the interference between SH fields requires the evaluation of second-order susceptibilities consisting of amplitude and phase. Whereas the former can be readily obtained by intensity measurements, the latter demands interferometric detection in general as reported earlier[31] and in this work. Although the employed spectral phase interferometry is robust because of its inline optical configuration,[31] it still requires somewhat sophisticated instrumentation. As an alternative, Kim et al.[31] used the ratio (denoted as R) between the minimum and maximum values of $I^{\parallel}(\theta)$ obtained from HS areas. Because R is a function of $\theta_S$ and $\Delta\varphi$ $(= \varphi_2 - \varphi_1)$, $\Delta\varphi$ can be determined with a set of R and $\theta_S$. Despite its simplicity, this method required multiple HS samples with various $\theta_S$ and mathematical simulations because $\Delta\varphi$ was not analytically given in terms of the experimental observables. As shown in Supplementary Note D, the **C**-matrix representation provides an analytical relation between $\Delta\varphi$ and R as follows:

$$\Delta\varphi = \left| \tan^{-1}\left[\frac{\tan 3(\theta_2^o - \theta_{max}^{\parallel})}{\sqrt{R}}\right] - \tan^{-1}\left[\frac{\tan 3(\theta_1^o - \theta_{max}^{\parallel})}{\sqrt{R}}\right] \right| \qquad (13)$$

, where $\theta^{\parallel}_{max}$ is the orientational angle for the maximum $I^{\parallel}$.

To validate the above equation, we compared its predictions with those of interferometric measurements. Figure 4a shows two pairs of interferograms obtained from each monolayer area of two $MoS_2/WS_2$ samples with $\theta_S = 15°$ and $30°$, respectively. They revealed the same phase difference of $61°$, which is independent of their stacking angle assuming negligible interlayer interaction[31]. In contrast, they differed in R and $\theta^{\parallel}_{max}$ values as shown in the $\theta$-dependence of $I^{\parallel}$ (Fig. 4b). In Fig. 4c, we show $|\Delta\varphi|$ calculated using Equation 13 for seven heterobilayer samples including the two in Fig. 4b: The phase difference between the two materials, 63.4 ± 4.3°, agreed with that determined by the interferometry. Clearly, $\Delta\varphi$ can be determined with a single heterobilayer sample of arbitrary stacking angle without interferometry.

The non-interferometric determination of $\Delta\varphi$ can also shed light on interlayer coupling in heterobilayer systems[43] because it is performed on the HS area whereas the interferometry probes 1L areas. We note that the superposition models proposed earlier[31, 32] and in this work neglected the interlayer coupling. Although it has been shown in this paper that the assumption was valid for a few material combinations (Fig. S8), it is not always true as shown recently[43]. Figure 4d shows that the two 1L areas of $WS_2/MoSe_2$ ($\theta_S = 30.1°$) undergo ~20° decrease in $\Delta\varphi$ with SH wavelength increasing from 400 to 450 nm. The polar intensity graphs in Fig. 4e also exhibited changes in their R and $\theta^{\parallel}_{max}$ values. Remarkably, $\Delta\varphi$ values determined by the two methods deviated from each other significantly except for 400 nm (Fig. 4f). These deviations indicate that the SH response from the HS area of $WS_2/MoSe_2$ is not equivalent to the sum of the contributions from the two 1L areas, which can be attributed to static or dynamic interlayer couplings[43]. The former may arise from the vdW interlayer bonding, which leads to the modification of electronic band structures[3, 48] and local intralayer chemical bonding[49]. The latter can be induced by photoexcitation and associated relaxation that transiently modulate the electric susceptibility of the system. In particular, the high peak power of the fundamental pulses used in a typical SHG spectroscopy readily allows two-photon excitations of excitons in the individual layers. It has been reported that interlayer charge transfer is highly efficient and occurs in the time scale of ~50 fs for TMD heterobilayers[13]. In this regard, we note that the SHG signals from $WS_2/MoSe_2$ are significantly modulated by photoinduced interlayer charge transfer[43], which is driven by the excitons generated in either layer.

***VdW-engineering of circularly polarized SHG.*** The generalized superposition model enables photonic design of vdW-stacked heterostructures that target SH fields with a specific polarization state. As a proof-of-concept demonstration, we chose to realize circularly polarized SH fields using linearly polarized fundamental beams. Circular polarization can be formed when two linearly polarized SH fields with equal amplitude and phase difference of 90° are superposed with their polarizations orthogonal to each other. Using two TMDs belonging to the $D^1_{3h}$ space group, for example, the orthogonality can be ensured by setting their stacking angle to 30°. However, the other two simultaneous conditions are not readily satisfied in general because the SHG amplitude and phase of the two materials vary distinctively as a function of SH wavelength[31]. As a generic solution depicted in Fig. 5a, we devised a ternary heterostructure consisting of two generation layers ($GL_1$ and $GL_2$) and a phase modulation layer (PML) that provides phase control to meet the phase requirement. Specifically, PML with normal dispersion will phase-delay the SH fields of $GL_1$ with respect to those of $GL_2$ when they reach a detector in the epi-detection scheme. As shown in Fig. 5b, the equal-amplitude requirement can also be met by using a homobilayer with a specific stacking angle ($\theta_{S,GL2}$) for a given SH wavelength.

Figure 5c shows one ternary heterostructure fabricated for circularly polarized SHG at $\lambda^{2\omega}$ = 420 nm. 1L $MoS_2$ and artificially stacked 2L $WS_2$ were selected respectively as $GL_1$ and $GL_2$ because of their extensive SHG information[31, 43]. Based on the design in Fig. 5a, an hBN slab with a thickness ($d_{hBN}$) of 15 nm was used as PML to increase the phase difference between the two GLs because their experimental $\Delta\varphi$ was much smaller than the required 90° for a wide wavelength range of interest (Fig. 5d): Neglecting multiple reflections, the phase delay induced by hBN is $\left(\frac{n^{\omega}_{hBN}}{\lambda^{\omega}} + \frac{n^{2\omega}_{hBN}}{\lambda^{2\omega}}\right) \cdot d_{hBN} \cdot 360°$, where $n^{\omega}_{hBN}$ and $n^{2\omega}_{hBN}$ are the refractive index of hBN at $\omega$ and $2\omega$, respectively. For example, at $\lambda^{2\omega}$ = 400 nm, where $n^{\omega}_{hBN}$ is 2.121 and $n^{2\omega}_{hBN}$ is 2.151[50], the phase delay by the PML was predicted to be 43°. Figure 5d shows that $\Delta\varphi$ indeed increased by 49° at 400 nm in the presence of PML and reached the targeted value of 90° near 420 nm. Note that the SHG signals from the PML was negligible (Fig. S10a). For the equal-amplitude requirement, $\theta_{S,GL2}$ was set as 42° (Fig. S10b). The relative orientation of $GL_2$ with respect to $GL_1$ was set to ensure the orthogonality condition depicted in Fig. 5b.

The $I^{\parallel}$ polar graphs of the two GL-only areas (Fig. 5e) revealed that their effective stacking angle ($\theta_{S,GL1-GL2}$) was 32.3 ± 0.2°, which indicates the two linearly polarized SHG fields are nearly orthogonal. The signals from the ternary heterostructure in Fig. 5f are close to what is expected from circular polarization. The slight ellipticity can be attributed to fabrication errors such as nonideal $\theta_{S,GL1-GL2}$ and some factors that were not incorporated into the design. Absorption of SH fields from $GL_1$ by $GL_2$ and multiple reflections need to be corrected for a precise design as will be discussed below. The interlayer interaction between two 1L $WS_2$ of $GL_2$ turned out to be also influential: The SHG intensity of $GL_2$ was 30% lower than the expectation calculated based on $\theta_{S,GL2}$ (Fig. S10a). A simulation using Equation 7 showed that no interaction within $GL_2$ would decrease the ellipticity significantly (Fig. S10c). The nature of interlayer interaction in $GL_2$ (twisted 2L $WS_2$) warrants a separate investigation.

***Further discussion - effects of reflection and absorption.*** We found that the modulation of SH amplitude and phase associated with optical absorption and reflection must be considered for high-precision measurements. As depicted in Fig. S11, SHG signals can undergo multiple reflections at two (1L) or three (2L) optical interfaces such as air-TMD, TMD-TMD and TMD-quartz interfaces. As shown in a previous study[32], the reflection at some of the interfaces can be significant because of the large refractive index of TMDs at the employed SH wavelengths. The optical absorption by TMD layers can also be nonnegligible at the SH wavelengths[43, 51]. To correct for these, we performed numerical simulations using the approach of Hsu et al[32]. Note that the heterobilayer case generates branching infinite series of light rays (Figs. S11c & S11d) unlike monolayers or homobilayers because of the additional TMD-TMD interface. However, the sum of the series can be given in analytic forms (Supplementary Note E). Some key results of the correction are as follows: Multiple reflections lead to a phase delay of -4.5° for $2\omega_{LO}$ ($\lambda^{2\omega}$ = 400 nm) with respect to the case where $2\omega_{LO}$ is completely reflected at the air-1L $MoS_2$ interface (Fig. S11a). Similarly, $2\omega_{sam}$ is also delayed by +4.1° with respect to the case where SHG generated at the center of 1L $MoS_2$ propagates backward without reflection (Fig. S11b). The presence of an adjacent layer also led to a noticeable phase shift: When $2\omega_{\mathrm{LO}}$ impinges on the $MoS_2/WS_2$ area, its phase is delayed by $-1.5°$ with respect to the $MoS_2$-only area (Fig. S11c). Similarly, $2\omega_{\mathrm{sam}}$ generated in the $MoS_2$-only area is delayed by $+2.1°$ when $WS_2$ is placed on the bottom (Fig. S11d).

When the corrections were made to $2\omega_{\mathrm{LO}}$ and $2\omega_{\mathrm{sam}}$, the SHG phase of $MoS_2$ ($\varphi_{\mathrm{Mo}}$) required for the evaluation of $\theta_{\mathrm{W}}^{\mathrm{o}}$ (Equations 11 and 12) was changed by $+3.6°$ for $\lambda^{2\omega} =$ 400 nm (see Fig. S12 for the effects of wavelength and stacking order). The validity of this correction can be appreciated by comparing the $\theta_{\mathrm{W}}^{\mathrm{o}}$ histograms with and without the correction (Fig. S13): Note that $\theta_{\mathrm{W}}^{\mathrm{o}}$ of the HS area approached that of 1L $WS_2$ within a sub-degree level (0.6°) after the correction.

## Conclusion

We developed DP-SPI for phase-resolved characterization of SHG signals from 2D TMDs and their heterocrystals. To validate the method, we demonstrated unprecedented interferometric orientational polarimetry that can distinguish and image antiparallel crystallographic domains of 1L TMDs and other materials of relevant symmetry. For comprehensive description of SHG in TMD heterobilayers, we devised a generalized superposition model in a matrix representation, which incorporates the material-dependent phase delay in addition to the SHG amplitude and crystallographic orientation of the two constituent monolayers into a **C** matrix characteristic of the material system. When combined with DP-SPI measurements, the superposition model enabled layer-resolved characterization of the heterobilayer. As an experimental verification, we performed spatial mapping of the crystallographic orientation for each monolayer of $MoS_2/WS_2$, which enabled direct stacking-angle imaging of heterobilayer areas. The matrix-based model also allowed determination of inter-material phase difference from conventional polarized SHG spectroscopy and nonlinear photonic design of a ternary heterostructure for circularly polarized SHG. Because of the experimental robustness of DP-SPI and generalized superposition model in efficient matrix representation, the approach demonstrated in this work can be applied to describing general heterostructures consisting of more than two monolayers, not limited to TMDs.

## Methods

***Preparation of samples.*** Monolayer $WS_2$ and $MoS_2$ samples were prepared by mechanical exfoliation[31] of bulk crystals (2D Semiconductors). Heterobilayer samples were assembled by dry-transferring a top layer exfoliated on a polydimethylsiloxane (PDMS) film onto a bottom layer supported on a quartz substrate[31]. For a specific stacking angle, the crystallographic orientation of each monolayer was determined prior to the dry transfer with polarized SHG measurements[31]. The positional and angular errors in the targeted transfer were less than 2 μm and 1°, respectively. Monolayer $WSe_2$ samples were grown by chemical vapor deposition on c-plane sapphire substrates[52].

***SHG spectroscopy.*** Conventional intensity measurements were performed with a home-built micro-SHG spectroscopy setup configured upon an optical microscope (Nikon, Ti-U)[31]. As a fundamental pulse, the linearly-polarized beam from a tunable Ti:sapphire laser (Coherent Inc., Chameleon) was focused on samples with a focal spot of 2.3 ± 0.2 μm in FWHM using a microscope objective (40×, numerical aperture = 0.60). The pulse duration and repetition rate were 140 fs and 80 MHz, respectively. The backscattered SHG signals were collected with the same objective and guided to a spectrometer equipped with a thermoelectrically cooled CCD detector (Andor Inc., DU971P). A linear polarizer was placed in front of the spectrometer to select the polarization of SHG signals. For polar graphs of SHG intensity, the orientation of samples was varied by rotating them about the surface normal using a rotational mount with a precision of 0.2°. Equivalently, the fundamental polarization may be rotated using a half waveplate with the sample and analyzing polarizer fixed in space[40, 53]. Note that resulting intensity polar graphs have 4-fold rotational symmetry.

***Dual-polarization spectral phase interferometry.*** The in-line spectral phase interferometry based on the above SHG spectroscopy setup was described elsewhere[31]. Briefly, local oscillator (LO) SHG pulses ($2\omega_{LO}$) were generated by focusing the fundamental beam at a 100 μm-thick z-cut α-quartz crystal (Fig. 1a), which could be rotated to vary the polarization of $2\omega_{LO}$. To avoid excessive time delay between LO and sample SHG pulses, a Cassegrain-type reflective objective (Edmund Optics, 52×, numerical aperture = 0.65) was used. The time delay was maintained in the range of 1 ~ 3 ps. Samples were raster-scanned using a piezo stage (Physik

Instrumente, P-545). For dual-polarization (DP) measurements, a Wollaston prism was placed in front of the spectrometer, and both components parallel and perpendicular to the fundamental's polarization were simultaneously obtained. For orientation imaging shown in Fig. 3, DP-SPI imaging was followed by DP-intensity imaging. Although DP interferograms contain intensity information, non-interferometric measurements provided higher precision. To prevent sample degradation, the average power density of the fundamental beam was maintained below 40 μW/μm$^2$ for interferometric measurements and below 100 μW/μm$^2$ for all others.

ASSOCIATED CONTENT

**Supporting Information.**

Polarized SHG intensity images of 1L $WSe_2$ crystals; Conventional and phase-resolved $\theta^o$ histograms; Fourier filtering of SHG interferograms; Raman and photoluminescence spectra of $MoS_2/WS_2$ in Fig. 2b; Polarized SHG analysis of $MoS_2/WS_2$ in Fig. 2b; Cross-configuration interferograms of $MoS_2/WS_2$ in Fig. 2b; Polarized SHG intensity and phase graphs of $MoS_2/WS_2$ in Fig. 2b; Polarized SHG intensity and phase graphs for various heterobilayers; Polarized SHG intensity and phase images of $MoS_2/WS_2$ in Fig. 3a; SHG intensity and polar graph of ternary heterostructure in Fig. 5c; Schematic for multiple reflections in $MoS_2$/quartz and $MoS_2/WS_2$/quartz; SHG phase shift induced by multiple reflections and absorptions at $MoS_2$-$WS_2$; Phase-shift correction on layer-resolved $\theta^o$ imaging; C matrix representation of SHG interference; General description of elliptically polarized SHG; Layer-resolved orientational imaging of heterobilayers; Non-interferometric determination of SHG phase; Effects of reflection and absorption on SHG. This material is available free of charge via the Internet at http://pubs.acs.org.

This work was submitted to a pre-preprint server:

Juseung Oh; Wontaek Kim; Gyouil Jeong; Yeri Lee; Jihun Kim; Hyeongjoon Kim; Hyeon Suk Shin; Sunmin Ryu. Dual-Polarization SHG Interferometry for Imaging Antiparallel Domains and Stacking Angles of 2D Heterocrystals. 2025, https://arxiv.org/abs/2505.21922 (May 28th, 2025)

AUTHOR INFORMATION

**Corresponding Author**

*E-mail: juseungoh@postech.ac.kr; sunryu@postech.ac.kr

**Author Contributions**

S.R. conceived the project. J.O. and S.R. designed the experiments. J.O., W.K., G.J., H.K. and H.S.S. prepared samples. J.O., W.K., Y.L. and J.K. performed the spectroscopy experiments and analyzed the data. J.O. and S.R. wrote the manuscript with contributions from all authors.

**ACKNOWLEDGMENTS**

This work was supported by the National Research Foundation of Korea (NRF-RS-2024-00336324, NRF-RS-2024-00411134, NRF-2021R1A6A1A10042944) and Samsung Electronics Co., Ltd (IO201215-08191-01). H.K. and H.S.S. acknowledge support from Institute for Basic Science (IBS-R036-D1).

NOTES

The authors declare no conflict of interest.

## Figures & Captions

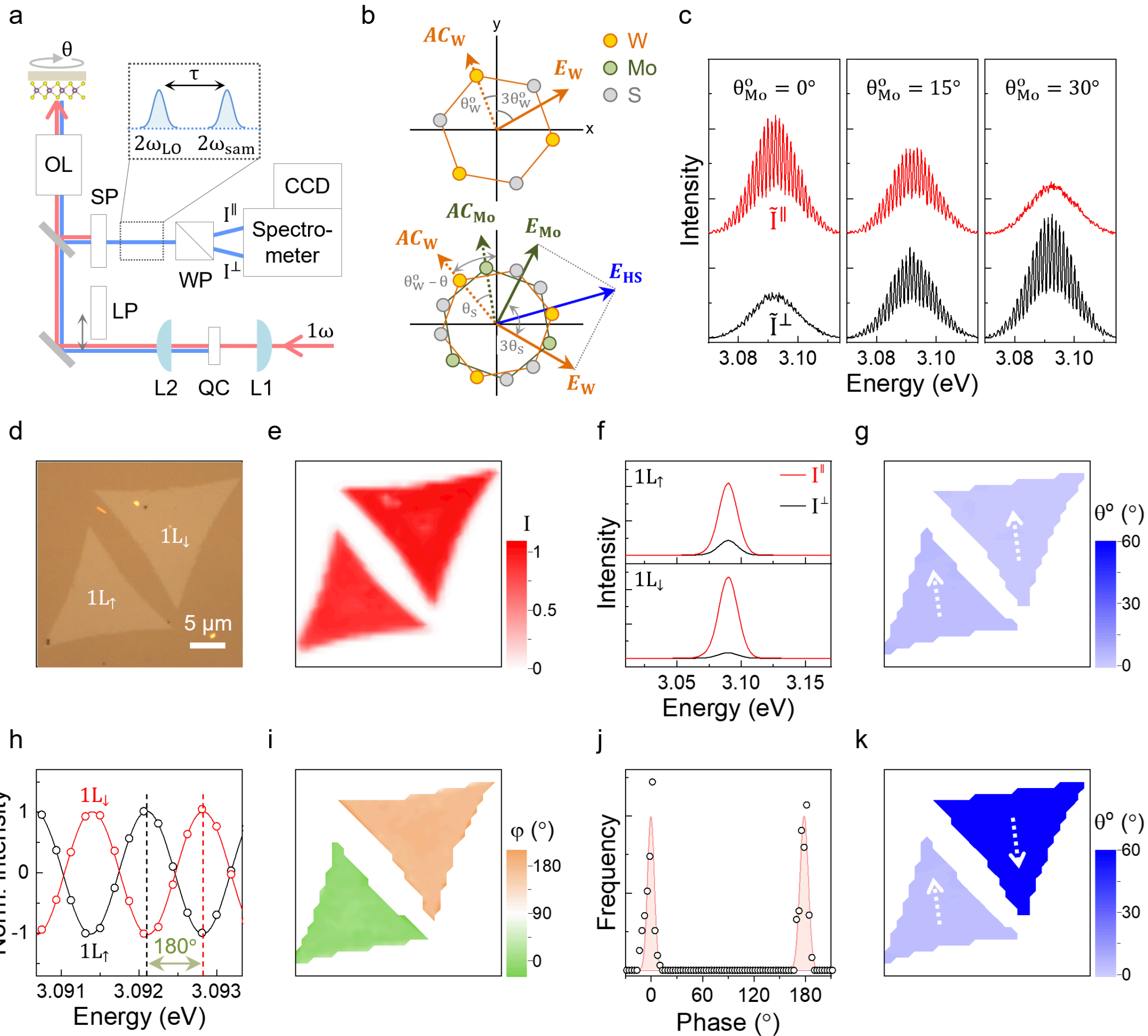

**Figure 1. Orientational polarimetry using DP-SPI.** (a) Optical layout of a DP-SPI setup: Convex lenses (L1 & L2), α-quartz crystal (QC), 52× objective lens (OL), long-pass filter (LP), short-pass filter (SP), Wollaston prism (WP), SHG signals of parallel & cross configuration ($I^{\parallel}$ & $I^{\perp}$), rotational angle of sample (θ), SH pulses of local oscillator and sample ($2\omega_{LO}$ & $2\omega_{sam}$) and delay time (τ) (see Methods for details). (b) Schemes for structure-polarization relation of 1L $WS_2$ (top) and $MoS_2/WS_2$ heterostructure, denoted as HS (bottom). $\theta^{o}_{W}$ is the angle between the $WS_2$ armchair direction ($\boldsymbol{AC_W}$) and the electric field of the fundamental beam (not shown) aligned along the +y axis. The resulting SH field ($\boldsymbol{E_W}$) forms $3\theta^{o}_{W}$ with the +y axis. The SH response from the heterostrucure ($\boldsymbol{E_{HS}}$) with stacking angle of $\theta_S$ can be described with the generalized superposition model given in the main text. Unlike $\theta^{o}$, θ is defined clockwise. (c) Polarized interferograms ($\tilde{I}^{\parallel}$ & $\tilde{I}^{\perp}$) of 1L $MoS_2$ obtained for three $\theta^{o}_{Mo}$

values. (d) Optical micrograph of two antiparallel CVD-grown 1L $WSe_2$ flakes ($1L_{\uparrow}$ & $1L_{\downarrow}$). (e & f) SHG intensity ($I = I^{\parallel} + I^{\perp}$) image (e) and polarized spectra (f) of $1L_{\uparrow}$ and $1L_{\downarrow}$. (g) Non-interferometric $\theta^{o}$ image of (d), where the dotted arrows mark the average values. (h) Representative DP interferograms of the two flakes, where a pair of vertical dashed lines denote their phase difference of 180°. (i ~ k) Phase image (i), phase histogram (j) and phase-resolved $\theta^{o}$ image (k) obtained from DP-SPI imaging of (d). A pair of dotted arrows in (k) correctly reveal antiparallel alignment of the two flakes.

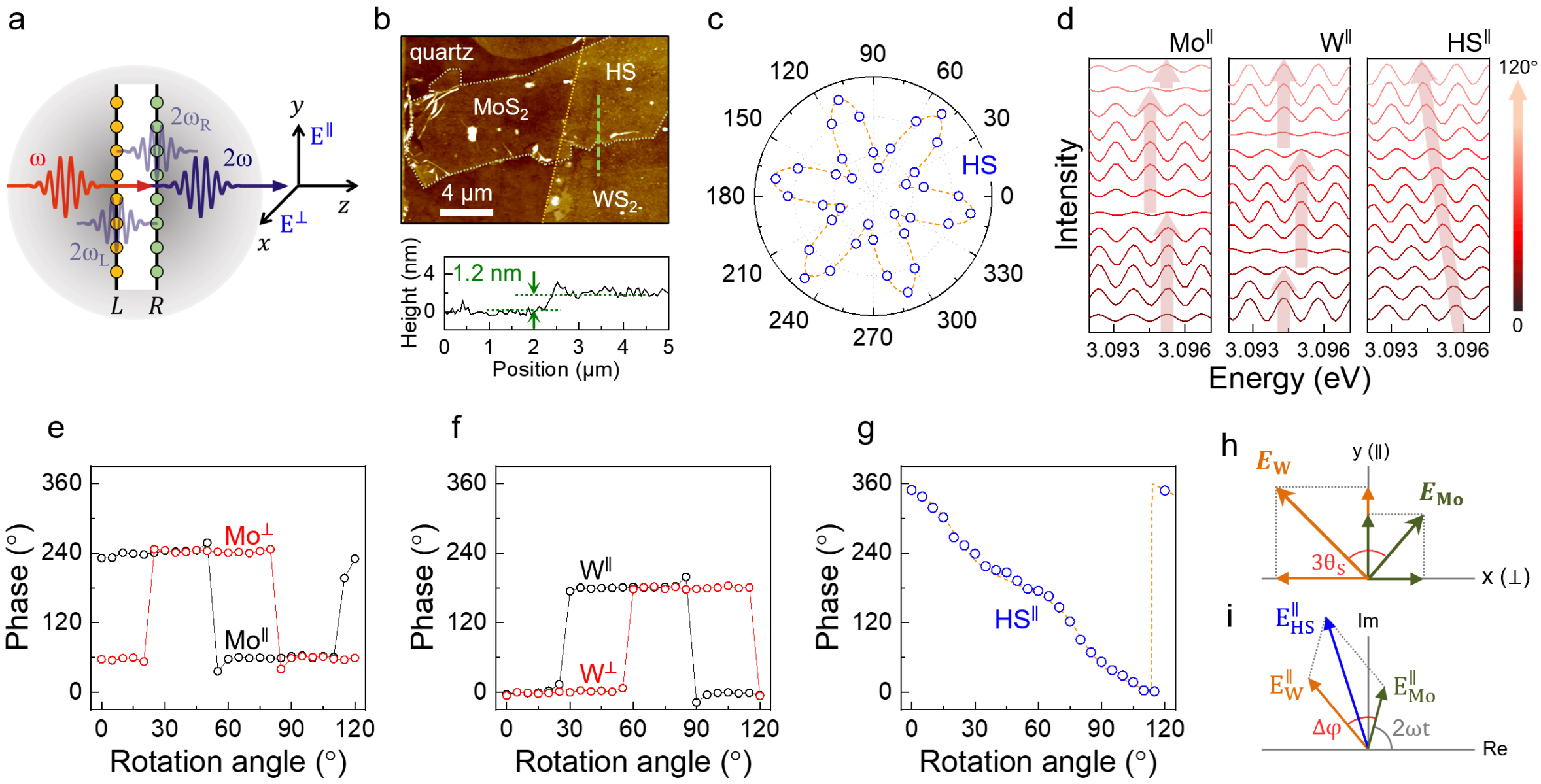

**Figure 2. Generalized SHG superposition model of heterobilayer.** (a) Schematic for superposition of two SH signals ($2\omega_L$ & $2\omega_R$) generated in heterobilayer consisting of two 1L 2D crystals (L & R) and DP detection of the overall signals ($2\omega$) propagating in the z direction. (b) AFM height image (top) and profile (bottom) of $MoS_2/WS_2$ heterostructure (HS) with $\theta_S = 32.9 \pm 0.2°$. The latter was obtained along the green dashed line in the former. (c) Polar graph of $I^{\parallel}$ obtained from the HS area. The orange dashed line represents the simulation based on the superposition model. (d) Polarized interferograms of 1L $MoS_2$ ($\tilde{I}^{\parallel}_{Mo}$), 1L $WS_2$ ($\tilde{I}^{\parallel}_{W}$), and HS ($\tilde{I}^{\parallel}_{HS}$) areas of (b), where $\theta$ was varied from 0° (bottom) to 120° (top) in step of 10°. (e ~ g) SHG phase ($\varphi$) of 1L $MoS_2$ (e) and 1L $WS_2$ (f) areas for the two polarization configurations ($\varphi^{\parallel}$ & $\varphi^{\perp}$), and HS area (g) for the parallel configuration. Phase values were referenced to that for $WS_2$ ($W^{\parallel}$) at $\theta =$ zero. The orange dashed line in (g) represents the simulation based on the superposition model. (h & i) Schemes for polarization decomposition of SH fields in laboratory coordinates (h) and superposition of two parallel SH fields on a complex plane (i).

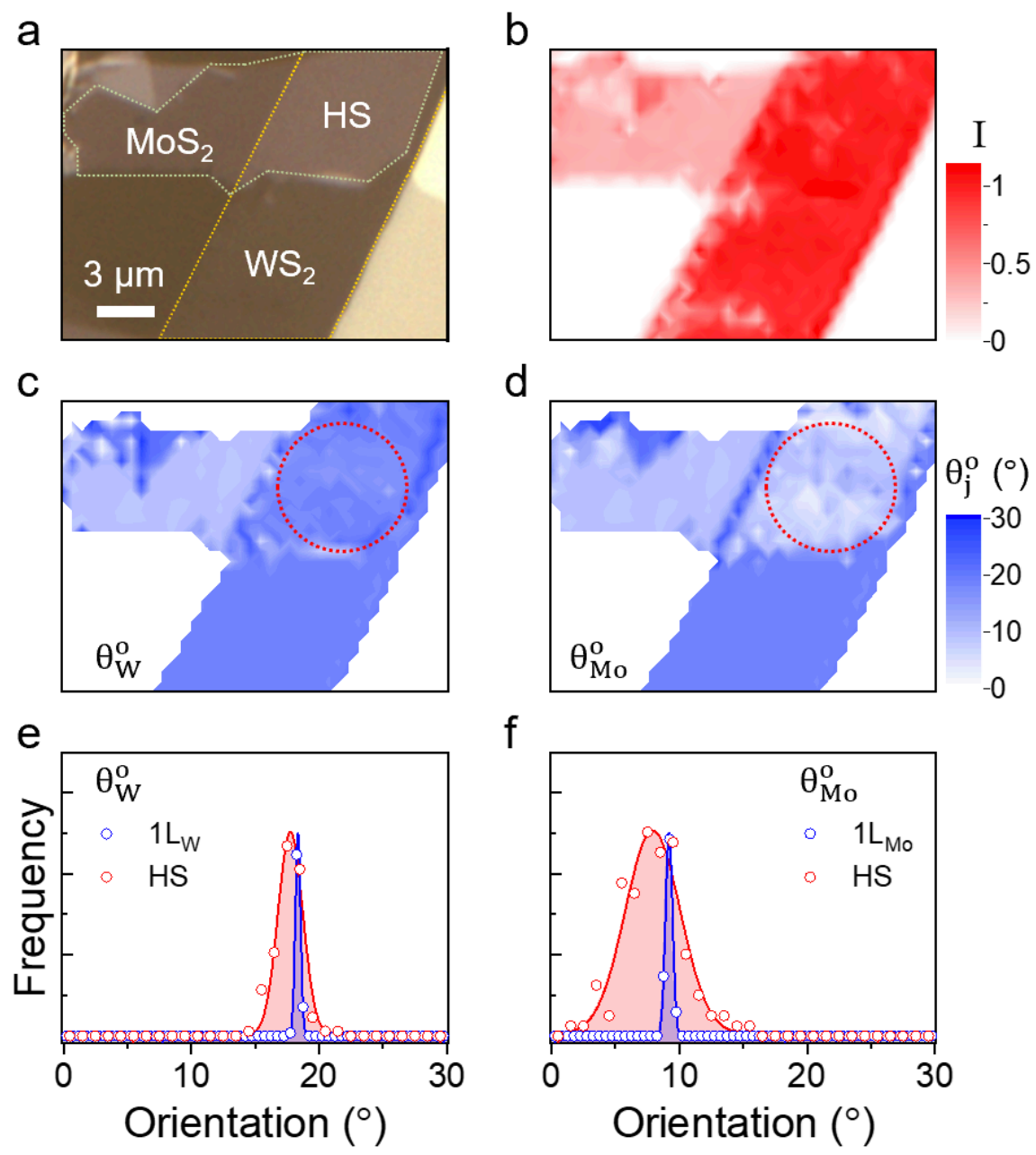

**Figure 3. Layer-resolved orientation imaging of heterobilayers.** (a) Optical micrograph of $MoS_2/WS_2$ heterobilayer shown in Fig. 2b. (b) SHG intensity ($I = I^{\parallel} + I^{\perp}$) image of (a). (c & d) Orientational angle images of $\theta^{o}_{W}$ (c) and $\theta^{o}_{Mo}$ (d). (e & f) Histograms of $\theta^{o}_{W}$ (e) and $\theta^{o}_{Mo}$ (f). For the HS area, the angle of each constituent layer was determined using the **C** matrix described in the main text and corrected for the phase shift induced by multiple reflections and absorptions (Fig. S13). The HS-area data in (e) and (f) were obtained within the dotted circles in (c) & (d), respectively.

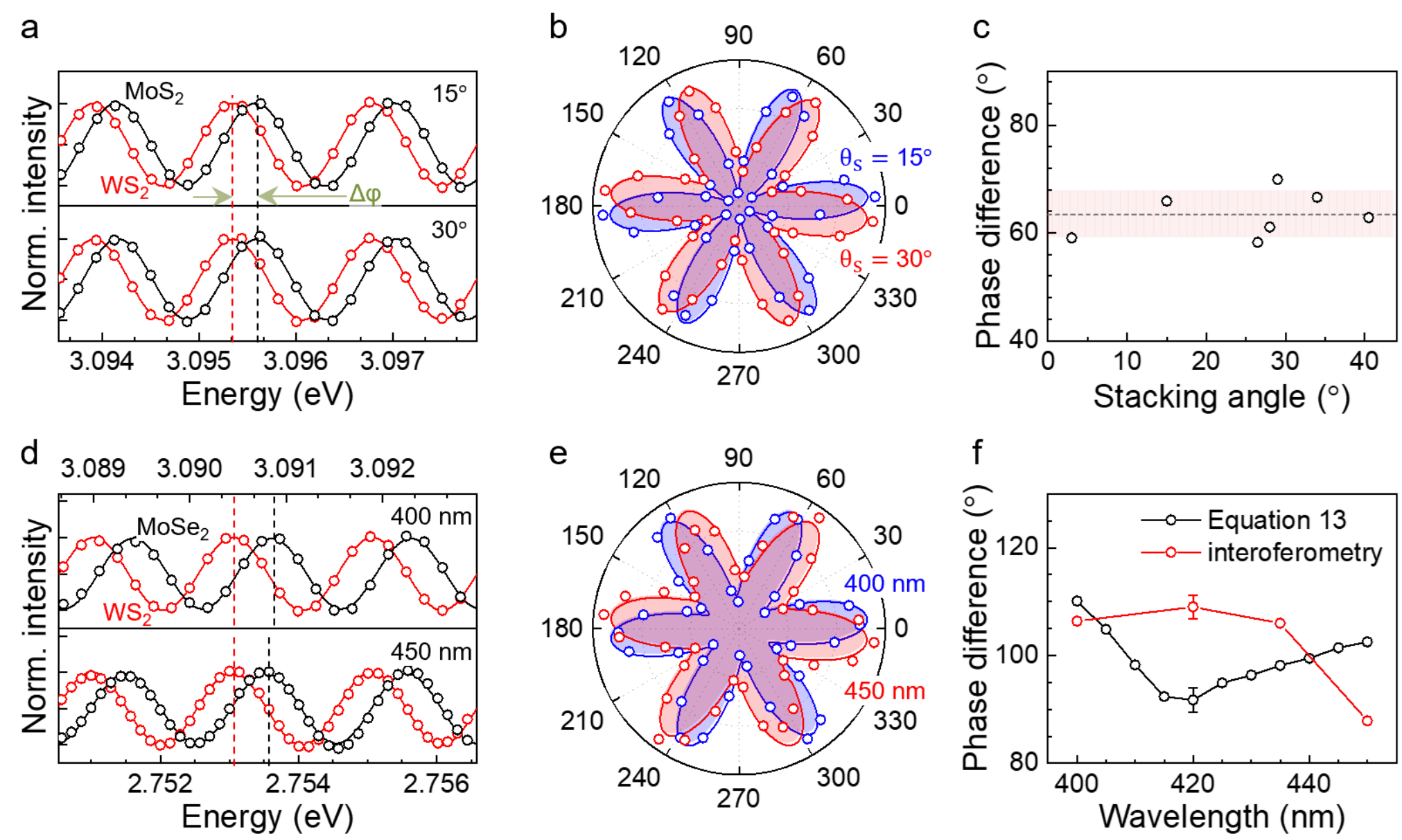

**Figure 4. Non-interferometric determination of inter-material phase difference.** (a) Interferograms of 1L areas of two $MoS_2/WS_2$ samples with $\theta_S = 15°$ and 30°. (b) Polar graphs of $I^{\parallel}$ obtained from the HS areas of the two samples in (a). (c) Inter-material phase difference ($\Delta\varphi$) determined for seven samples with various $\theta_S$ using their R values (see the main text for details). Dashed line and shade denote the average and standard deviation, respectively. (d) Interferograms obtained for two SH wavelengths (400 and 450 nm) from a $WS_2/MoSe_2$ sample with $\theta_S = 30°$. (e) Polar graphs of $I^{\parallel}$ obtained for the two wavelengths from the HS area of the sample in (d). (f) $\Delta\varphi$ determined for the sample in (d) by two methods for various wavelengths. Pairs of vertical dashed lines in (a) and (d) denote $\Delta\varphi$.

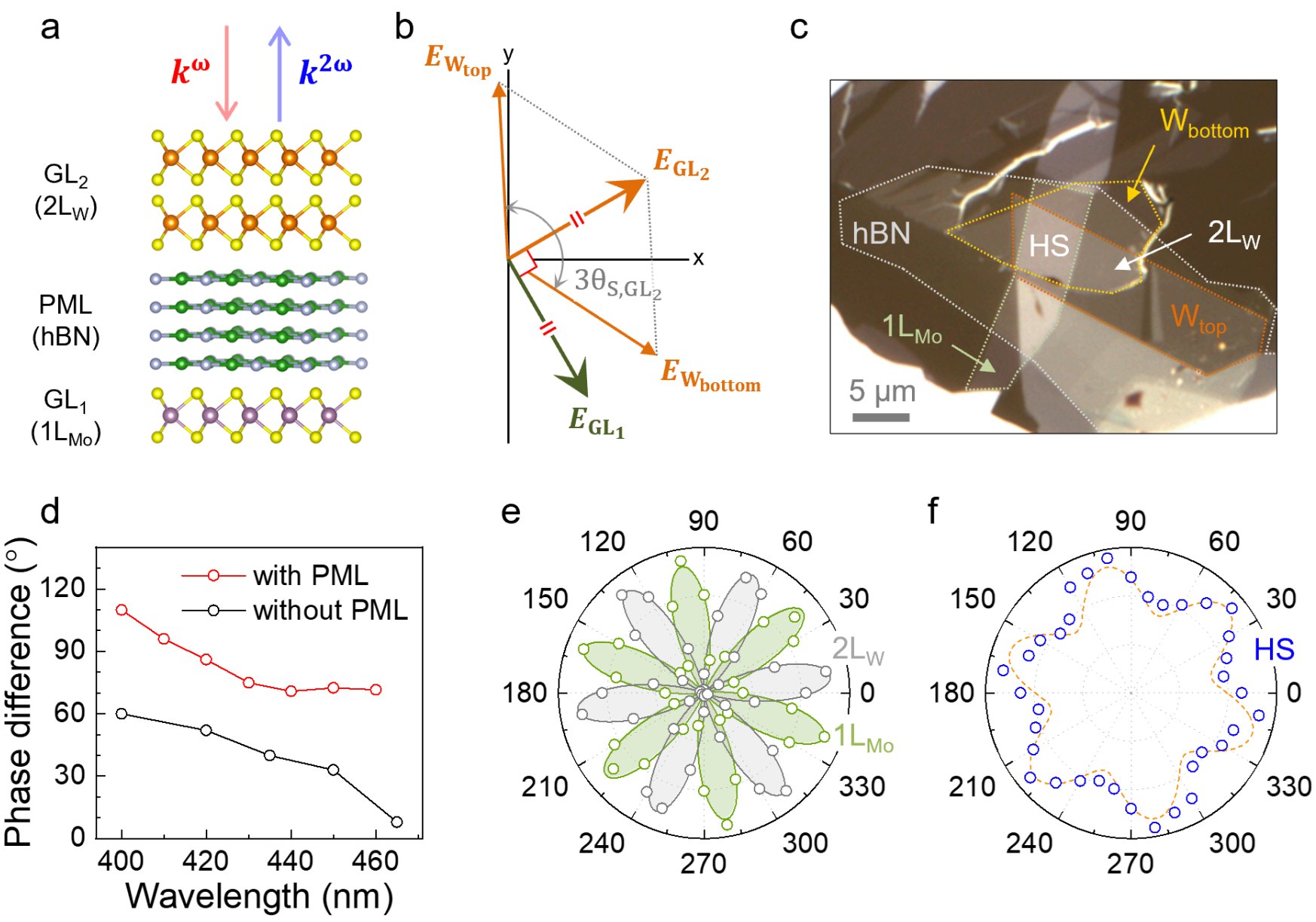

**Figure 5. Photonic design of ternary heterostructures for circularly polarized SHG.** (a) Ternary heterostructure consisting of two generation layers (1L $MoS_2$ as $GL_1$ and 2L $WS_2$ as $GL_2$) and a phase modulation layer (PML, 15 nm-thick hBN). SHG beam propagates from bottom to top ($\boldsymbol{k}^{2\omega}$) in opposite to the fundamental beam ($\boldsymbol{k}^{\omega}$). (b) Schematic for the equal-amplitude and orthogonality requirements. The stacking angle of 2L $WS_2$ ($\theta_{S,GL2}$) and that between $GL_1$ and $GL_2$ ($\theta_{S,GL1-GL2}$) were selected to ensure the equal SH amplitudes ($|\boldsymbol{E}_{\mathbf{GL1}}| = |\boldsymbol{E}_{\mathbf{GL2}}|$) and polarization orthogonality ($\boldsymbol{E}_{\mathbf{GL1}} \perp \boldsymbol{E}_{\mathbf{GL2}}$), as described in the main text. (c) Optical micrograph of an artificially stacked ternary heterostructure constructed to the design. (d) SHG phase difference between $GL_1$ and $GL_2$ with and without PML. (e) Polar graph of $I^{\parallel}$ obtained from of $GL_1$ and $GL_2$-only areas, which showed that $\theta_{S,GL1-GL2}$ is 32.3 ± 0.2°. (f) Polar graph of $I^{\parallel}$ obtained from the ternary HS area of (c). The orange dashed line represents the simulation based on the superposition model.

**TOC Graphic**

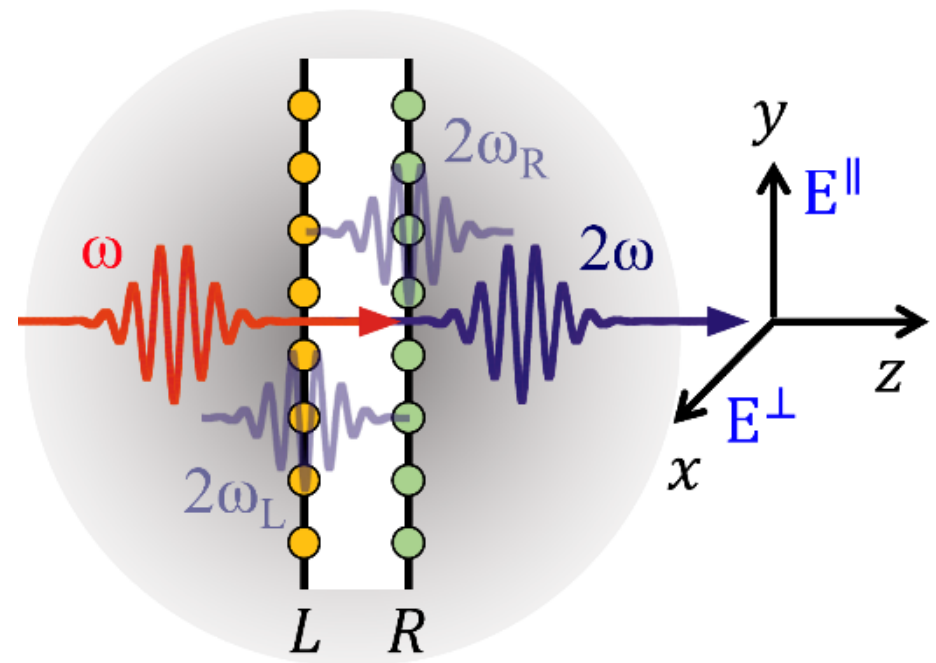